\documentclass[aps,prl,twocolumn,superscriptaddress]{revtex4}
\usepackage{graphicx}
\usepackage{amssymb}
\usepackage{amsmath}
\usepackage{graphicx}
\usepackage{epsfig}
\usepackage{array}
\usepackage{color}
\usepackage{multirow}
\usepackage{ulem}

\begin{document}

\title{Flat bands promoted by Hund's rule coupling in the candidate
double-layer high-temperature superconductor La$_3$Ni$_2$O$_7$}

\author{Yingying Cao}
\affiliation{Beijing National Laboratory for Condensed Matter Physics and
Institute of Physics, Chinese Academy of Sciences, Beijing 100190, China}
\affiliation{University of Chinese Academy of Sciences, Beijing 100049, China}
\author{Yi-feng Yang}
\email[]{yifeng@iphy.ac.cn}
\affiliation{Beijing National Laboratory for Condensed Matter Physics and
Institute of Physics, Chinese Academy of Sciences, Beijing 100190, China}
\affiliation{University of Chinese Academy of Sciences, Beijing 100049, China}
\affiliation{Songshan Lake Materials Laboratory, Dongguan, Guangdong 523808, China}

\date{\today}

\begin{abstract}
We report strongly correlated electronic band structure calculations of the recently discovered double-layer high-temperature superconductor La$_3$Ni$_2$O$_7$ under pressure. Our calculations reveal dual nature of Ni-$d$ electrons with almost localized $d_{z^2}$ orbitals due to onsite Coulomb repulsion and very flat hybridization bands of Ni-$d_{x^2-y^2}$ and Ni-$d_{z^2}$ quasiparticles near the Fermi energy. We find that the quasiparticle effective mass are greatly enhanced by the Hund's rule coupling and their lifetimes are inversely proportional to the temperature, which explains the experimentally observed strange metal behavior in the normal state. We also find strong antiferromagnetic spin correlations of Ni-$d$ electrons, which may provide the pairing force of quasiparticles for the high-temperature superconductivity. These give a potential explanation of two key observations in experiment and connect the superconducting La$_3$Ni$_2$O$_7$ with cuprate high-temperature superconductors. The presence of flat bands and the interplay of orbital-selective Mott, Hund, and Kondo physics make La$_3$Ni$_2$O$_7$ a unique platform for exploring rich emergent quantum many-body phenomena in the future.
\end{abstract}

\maketitle

Recent discovery of possible high-temperature superconductivity in the pressurized Ruddlesden-Popper perovskite La$_3$Ni$_2$O$_7$ \cite{Sun2023} has stimulated intensive investigations on its fundamental electronic structures and potential pairing mechanism \cite{Sakakibara2023, Luo2023, Yang2023, Zhang2023, Lechermann2023, Shilenko2023, Christiansson2023, Shen2023, Gu2023, Liu2023}. The Ni cation is expected to have a valence of +2.5. First-principles calculations yield an almost fully filled Ni-$d_{z^2}$ bonding state and quarter-filled Ni-$d_{x^2-y^2}$ bands due to the double-layer structure \cite{Sun2023,Nakata2017PRB}, suggesting a weak correlation picture different from the cuprate high-temperature superconductors and the infinite-layer nickelate superconductors, in which the transition metal ions (Cu$^{2+}$ or Ni$^{1+}$) have a $d^9$ configuration with (almost) half-filled and localized $d_{x^2-y^2}$ orbital due to the Mott localization \cite{Lee2006RMP, Lee2004PRB, Li2019Nature, Zhang2020PRB,Yang2022FP}. Questions arise on how the weakly correlated bands in La$_3$Ni$_2$O$_7$ can support the experimentally observed high-temperature superconductivity around 80 K, and what electronic features are of primary importance for its Cooper pairing and the strange metal behavior observed in the normal state \cite{Sun2023}.

In this work, we show that the above weak correlation picture is incorrect for superconducting La$_3$Ni$_2$O$_7$ by performing a systematic investigation of its strongly correlated electronic band structures using the density functional theory (DFT) \cite{2014WIEN2k, WIEN2k} combined with the dynamical mean-field theory (DMFT) \cite{Georges1996RMP, Anisimov1997JPCM, Lichtenstein1998PRB, Kotliar2006RMP, Held2008JPCM, Haule2010PRB}. Our calculations reveal itinerant-localized duality of strongly correlated $d_{z^2}$ electrons with both quasiparticle bands and local moment fluctuations. The $d_{x^2-y^2}$ bands are also strongly renormalized near the Fermi energy by hybridizing with the $d_{z^2}$ bands. These differ distinctively from pure DFT prediction but resemble those in orbital-selective Mott systems or Kondo lattice systems \cite{Anisimov2002EPJB,Vojta2010JLTP,Shim2007S,Xu2022npjQM}. The Hund's rule coupling is found to compete with the hybridization, causing very flat bands near the Fermi energy. Our self-energy analyses yield the quasiparticle lifetimes inversely proportional to the temperature, which explains the strange metal behavior observed in the normal state resistivity \cite{Sun2023}. The static local spin susceptibility follows typical Curie-Weiss behavior, confirming the presence of $d_{z^2}$ local moments, and gives a large antiferromagnetic coupling possibly from their inter-layer superexchange interaction through the apical O-$p_z$ orbitals. These imply that the Cooper pairing might be primarily induced by antiferromagnetic spin fluctuations of the almost localized $d_{z^2}$ electrons and become coherent through hybridization (and self-doping) with $d_{x^2-y^2}$ electrons, thus connecting the double-layer La$_3$Ni$_2$O$_7$ to the cuprate high-temperature superconductors \cite{Lee2006RMP}. Our work explains two key experimental observations and provides a natural framework for exploring the physics of La$_3$Ni$_2$O$_7$.

\begin{figure}[t]
    \begin{center}
        \includegraphics[width=0.48\textwidth]{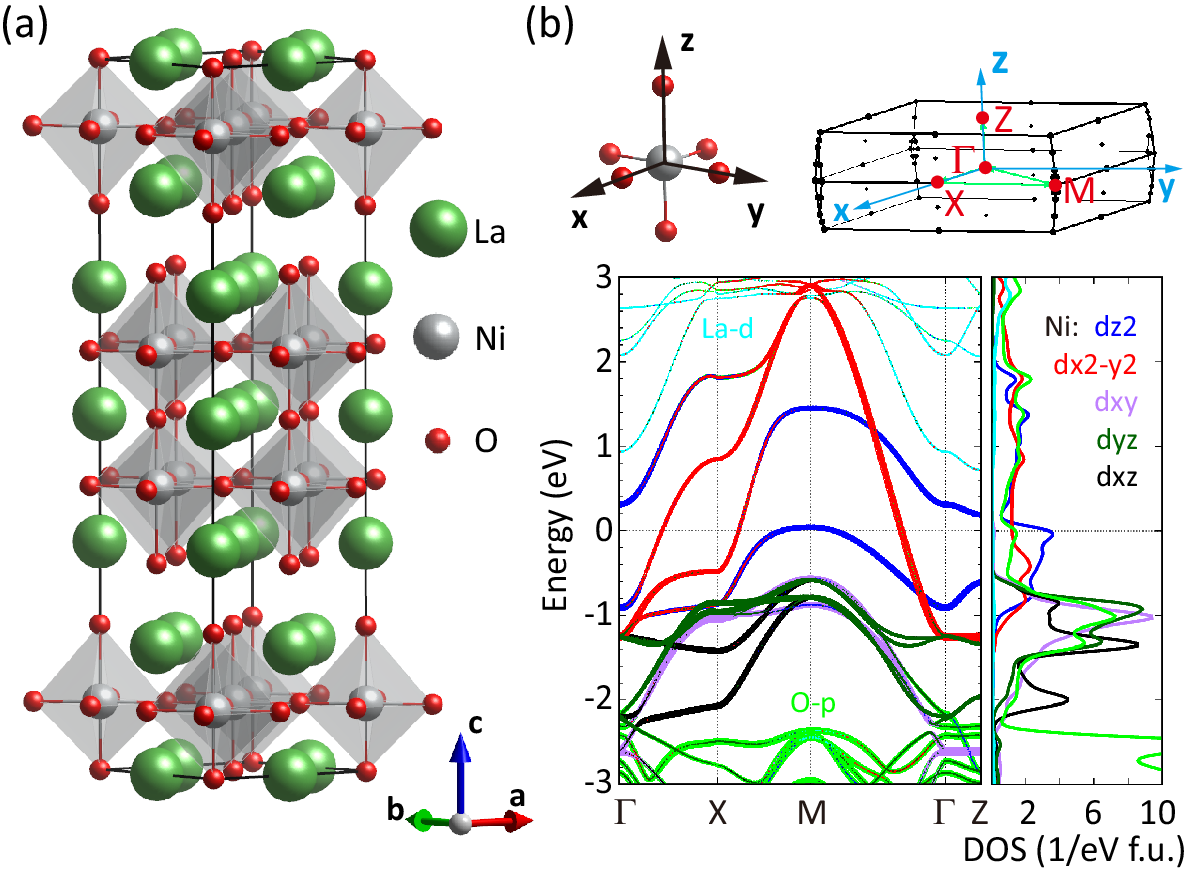}
        \caption{(a) High-pressure crystal structure of the candidate high-temperature superconductor La$_3$Ni$_2$O$_7$ visualized by the VESTA \cite{VESTA}. (b) Orbital-projected  band structures and densities of states for Ni-$d$, O-$p$, and La-$d$ calculated by DFT at 29.5 GPa, with optimized structural parameters based on experiment \cite{Sun2023}. The paths are shown as green lines in the schematic diagram of the Brillouin zone plotted using the Xcrysden \cite{Xcrysden}, where the local $z$-axis points along the $c$ direction and the $x$- and $y$-axes along the Ni-O bonds in the $ab$ plane. The red points mark the high symmetry points in the Brillouin zone.}
        \label{fig1}
    \end{center}
\end{figure}

The superconducting La$_3$Ni$_2$O$_7$ has an orthorhombic structure with the space group $Fmmm$. As shown in Fig. \ref{fig1}(a), each Ni ion is surrounded by six O ions to form an octahedron, producing a double-layer structure with two Ni ions sharing an apical O. For later comparison, Fig. \ref{fig1}(b) gives the DFT band structures calculated using WIEN2k \cite{WIEN2k, Perdew1996PRL}. The overall results are similar as reported in previous DFT calculations \cite{Sun2023}. We obtain a metal with three bands crossing the Fermi level. These bands are primarily from Ni-$d_{z^2}$ and Ni-$d_{x^2-y^2}$ orbitals but contain substantial contributions from O-$p$ orbitals. The Ni-$d_{x^2-y^2}$ bands (red) have a large bandwidth of about 4 eV and are strongly hybridized with $d_{z^2}$. The $d_{z^2}$ orbitals split into bonding and antibonding states due to strong inter-layer coupling through the apical O-$p_z$ orbital, giving rise to two bands (blue) with typical bandwidth of 1 eV separated by about 1.4 eV around M. The bonding state is almost fully occupied, while the antibonding state is  above the Fermi energy and considered irrelevant in previous studies, which leads to an effective low-energy model with an almost fully-filled $d_{z^2}$ bonding band and two quarter-filled $d_{x^2-y^2}$ bands. We will see that this DFT picture is insufficient due to strong electronic correlations of Ni-$d$ electrons. In particular, the doubly-occupied $d_{z^2}$ bonding state will be replaced by almost localized $d_{z^2}$ electrons on each Ni ion and flat quasiparticle bands due to their hybridization and Hund's rule coupling with more itinerant $d_{x^2-y^2}$ electrons. The pressurized La$_3$Ni$_2$O$_7$ is therefore a unique system with peculiar orbital-selective Mott, Hund, and Kondo physics.

\begin{figure}[t]
    \begin{center}
        \includegraphics[width=0.48\textwidth]{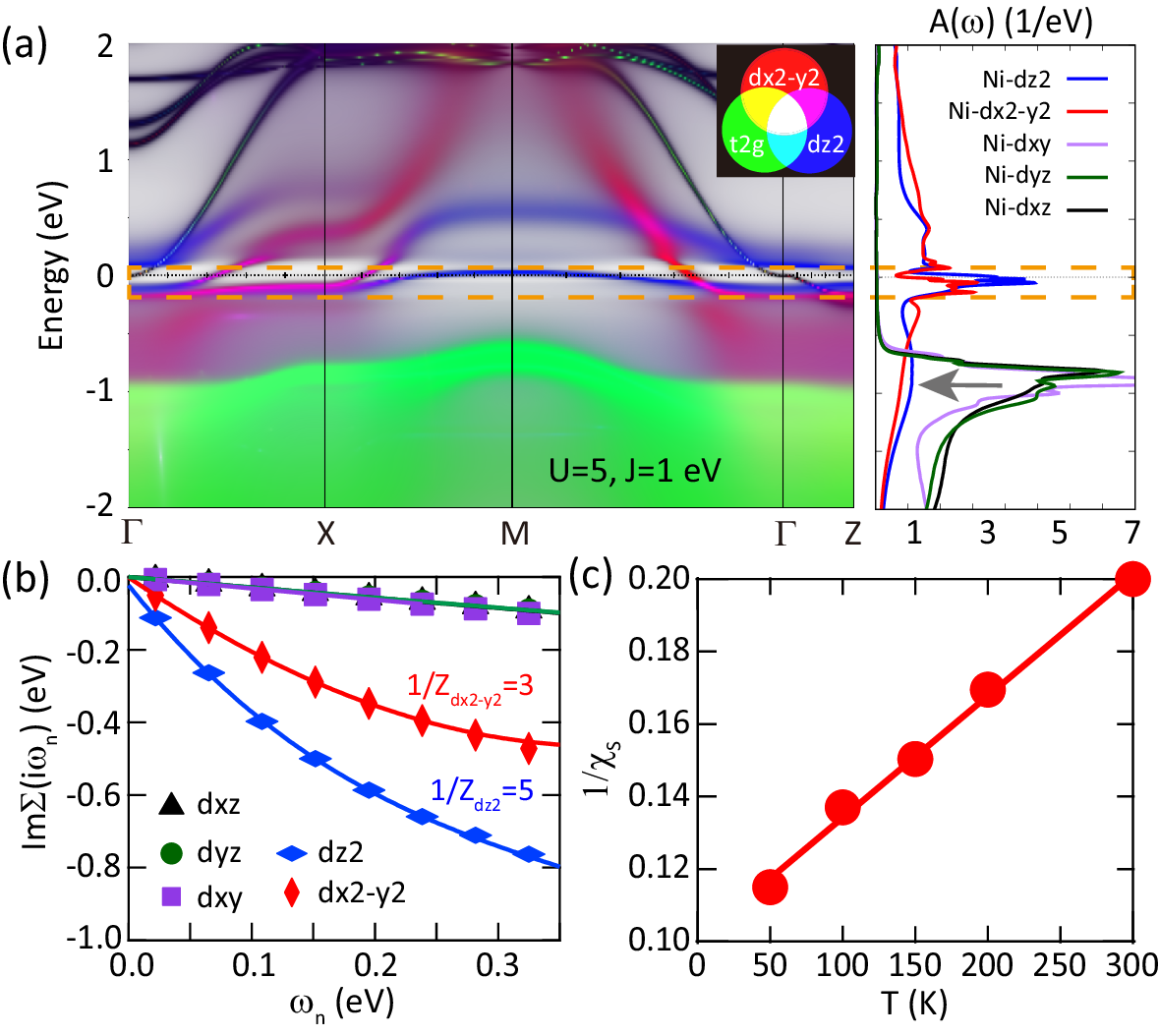}
        \caption{(a) Orbital-projected spectral function of Ni-$d$ electrons calculated using DFT+DMFT for $U=5$ eV and $J=1$ eV at 80 K. The red color is assigned to $d_{x^2-y^2}$, blue to $d_{z^2}$, and green to $t_{2g}$. Also shown are the corresponding momentum-integrated densities of states, with the three orbital components of $t_{2g}$ being seprated. The orange box highlights the peak from the flat quasiparticle bands of the almost localized $d_{z^2}$ orbitals. The gray arrow highlights the lower Hubbard band of Ni-$d_{z^2}$. The $d_{x^2-y^2}$ orbital shows similar features in the density of states due to its strong hybridization with the $d_{z^2}$ orbital. (b) Imaginary-part of the orbital-dependent self-energy for $U = 5$ eV and $J = 1$ eV at 80 K. The solid lines are fittings to the data using the fourth-order polynomial, giving an inverse renormalization factor of 5 for $d_{z^2}$ and 3 for $d_{x^2-y^2}$ at the Fermi energy. (c) Temperature dependence of the inverse static local spin susceptibility $1/\chi_s$. The solid line marks a linear-in-temperature fit according to the Curie-Weiss law.}
        \label{fig2}
    \end{center}
\end{figure}

Figure \ref{fig2}(a) plots our DFT+DMFT spectral function calculated at 80 K with the onsite Coulomb repulsion $U = 5$ eV and the Hund's rule coupling $J = 1$ eV on the Ni-$d$ electrons. The values of $U$ and $J$ are chosen according to previous study based on constrained random-phase approximation \cite{Sasioglu2011PRB}. We take the hybridization expansion continuous-time quantum Monte Carlo as the impurity solver \cite{Haule2007PRB} and the nominal scheme for double counting. To achieve high accuracy, more than $3 \times 10^7$ Monte Carlo steps are performed on each of the 56 processors. The spectral function is obtained by using the maximum entropy method for analytic continuation of the self-energy \cite{Jarrell1996PR}. Compared to DFT, the bonding-antibonding splitting is greatly reduced. The Ni-$d_{z^2}$ quasiparticle band around M is also strongly renormalized and becomes very flat near the Fermi energy. Its bandwidth is only about 0.2 eV, a factor of five narrower than the DFT prediction. The exact value of the band renormalization is derived in Fig. \ref{fig2}(b) from $Z^{-1} = 1- \partial$Im$\Sigma(i\omega)/\partial\omega|_{\omega\to0^+}$ by fitting the lowest six points of the imaginary part of the orbital-dependent self-energy with a fourth-order polynomial \cite{Mravlje2011PRL}. We find a mass enhancement of 5 for the Ni-$d_{z^2}$ quasiparticles, consistent with its reduced bandwidth from about 1 to 0.2 eV.

Interestingly, the Ni-$d_{x^2-y^2}$ electrons also have a large mass enhancement of 3. This seems odd because the $d_{x^2-y^2}$ bands are quarter-filled and far away from the Mott regime. We notice that in the band plot Fig. \ref{fig2}(a), this large renormalization only occurs near the Fermi energy, where $d_{x^2-y^2}$ hybridizes strongly with $d_{z^2}$ and produces some quite flat bands ($\Gamma$-X) with both orbital characters. Away from the Fermi energy, the $d_{x^2-y^2}$ bands are less renormalized, with the top at about 2.5 eV compared to 3 eV in DFT. Thus, the large renormalization of $d_{x^2-y^2}$ electrons near the Fermi energy mainly originates from their hybridization with $d_{z^2}$, as is also confirmed from the similar peak structures in their densities of states. This peculiar hybridization feature is close to those in heavy fermion systems \cite{Shim2007S,Xu2022npjQM}, where the conduction bands are only bent near the Fermi energy due to the Kondo hybridization with localized orbitals. On the other hand, because the atomic wave functions of Ni-$d_{z^2}$ and Ni-$d_{x^2-y^2}$ orbitals are orthogonal, their hybridization in La$_3$Ni$_2$O$_7$ should mostly occur from the hopping between neighboring Ni sites through the O-$p$ orbitals.

Connections to the Kondo or orbital-selective Mott physics are further supported by the almost localized behavior of $d_{z^2}$ electrons. Different from previous studies \cite{Sun2023,Zhang2023, Lechermann2023, Shilenko2023, Christiansson2023}, we see in Fig. \ref{fig2}(a) the blurred lower Hubbard band below the Fermi energy, as also marked by the arrow at around -1 eV in the orbital-resolved densities of states. This localized spectral weight indicates the dual nature  of strongly correlated Ni-$d_{z^2}$ electrons. To confirm it, we further calculate the static local spin susceptibility $\chi_s$ of Ni-$d$ electrons as a function of the temperature. As shown in Fig. \ref{fig2}(c), its inverse follows nicely the Curie-Weiss law (the straight line), which confirms the presence of well-defined local moments on Ni ions. The linear extrapolation gives a large Weiss temperature of approximately $-300$ K, indicating strong antiferromagnetic spin correlations among Ni-$d$ electrons. Since the  Hund's rule coupling between $d_{z^2}$ and $d_{x^2-y^2}$ electrons is ferromagnetic, the large antiferromagnetic correlations should mostly arise from the inter-layer superexchange coupling between two Ni-$d_{z^2}$ spins through the apical O-$p_z$ orbital. In cuprates, the superexchange interaction between localized Cu-$d_{x^2-y^2}$ spins gives the basic magnetic energy scale for the Cooper pairing. One may speculate that superexchange mechanism also plays a similar role here in superconducting La$_3$Ni$_2$O$_7$.

\begin{figure}[t]
    \begin{center}
        \includegraphics[width=0.48\textwidth]{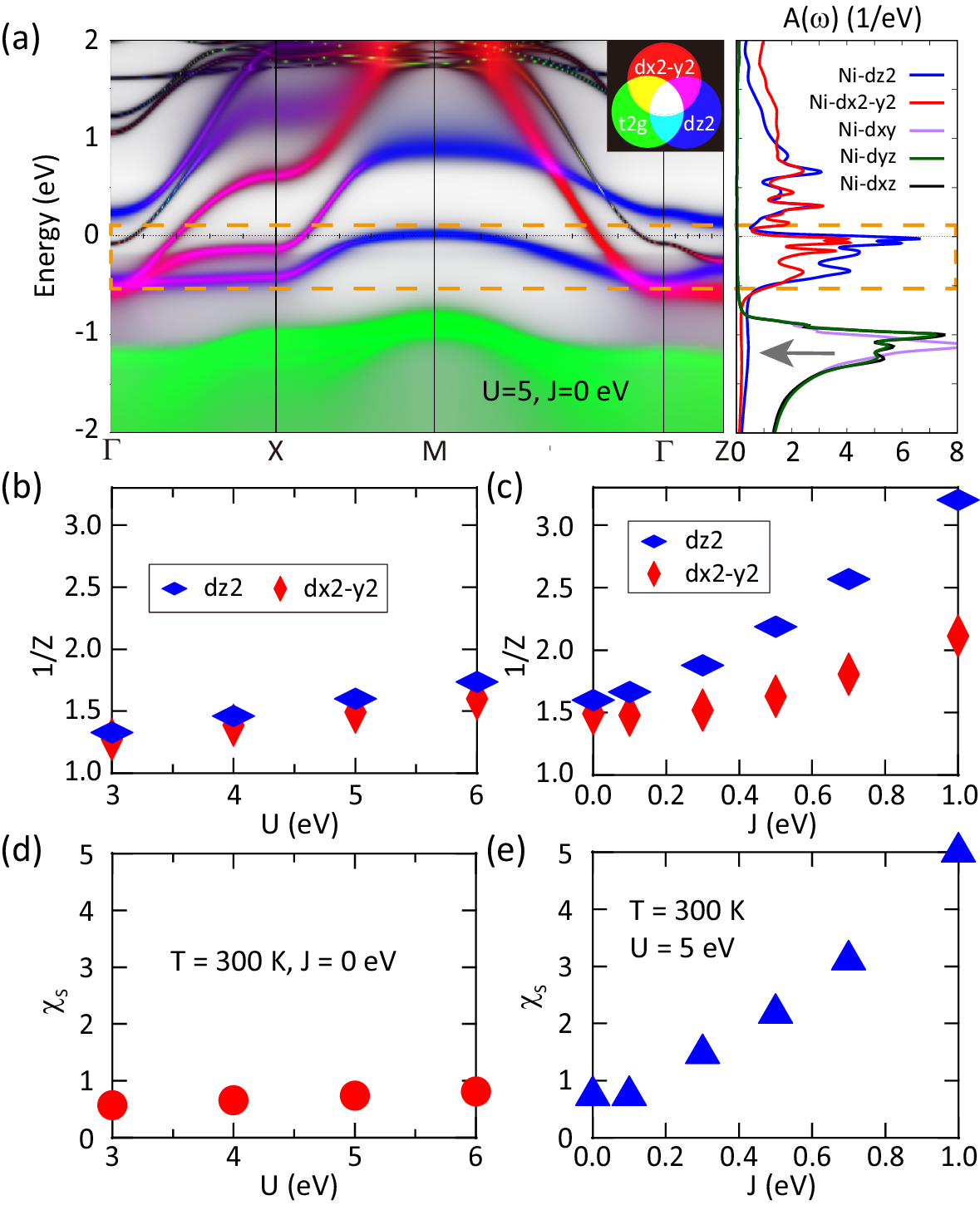}
        \caption{(a) Orbital-projected spectral function and corresponding densities of states of Ni-$d$ electrons calculated using DFT+DMFT for $U=5$ eV and $J=0$ eV at 80 K, plotted as in Fig. \ref{fig2}(a) showing reduced band renormalization. (b,c) Inverse renormalization factor of Ni-$d_{z^2}$ and Ni-$d_{x^2-y^2}$ orbitals extracted from the self-energy at 300 K as functions of the Coulomb repulsion $U$ for $J=0$ and of the Hund's rule coupling $J$ for $U=5$ eV, respectively. (d,e) The static local spin susceptibility $\chi_s$ as functions of $U$ for $J=0$ and of $J$ for $U=5$ eV, respectively.}
        \label{fig3}
    \end{center}
\end{figure}

The Hund's rule coupling between $d_{z^2}$ and $d_{x^2-y^2}$ orbitals could be also important. To see its effect, we compare in Fig. \ref{fig3}(a) the calculated spectral function for $U=5$ eV and $J=0$. Quite unexpectedly, the Ni-$d_{z^2}$ bands become much broader (0.7 eV), and the $d_{x^2-y^2}$ bandwidth increases to about 3 eV, which is close to the DFT value. It seems that the Hund's rule coupling plays an important role in the renormalization. For clarification, we plot the renormalization factor $1/Z$ as functions of $U$ and $J$ at 300 K in Figs. \ref{fig3}(b,c). Contrary to the usual expectation, for $J=0$, $1/Z$ varies only slightly from 1.3 to 1.7 as $U$ increases from 3 to 6 eV and is almost the same for two orbitals. But if we fix $U=5$ eV and increase $J$ from 0 to 1 eV, the inverse renormalization factor increases rapidly from about 1.6 to 3.2 for $d_{z^2}$ and 1.5 to 2.1 for $d_{x^2-y^2}$. Thus, the Hund's rule coupling greatly enhances the quasiparticle renormalization and promotes the flat bands near the Fermi energy. It also induces large difference in their orbital dependence. As a result, the Ni-$d_{z^2}$ band becomes noticeably more flattened with increasingly $J$. This is quite unusual, given the small variation of the Hund's rule coupling compared to the Hubbard interaction $U$. It could be that the ferromagnetic Hund's rule coupling tends to compete with the hybridization, thus reducing the effective Kondo energy scale that determines the quasiparticle bandwidth. Figures \ref{fig3}(d,e) confirm this idea by showing the static local spin susceptibility as functions of $U$ and $J$, which grows significantly at large $J$, reflecting enhanced local moment fluctuations due to the suppression of hybridization.

\begin{figure}[t]
    \begin{center}
        \includegraphics[width=0.45\textwidth]{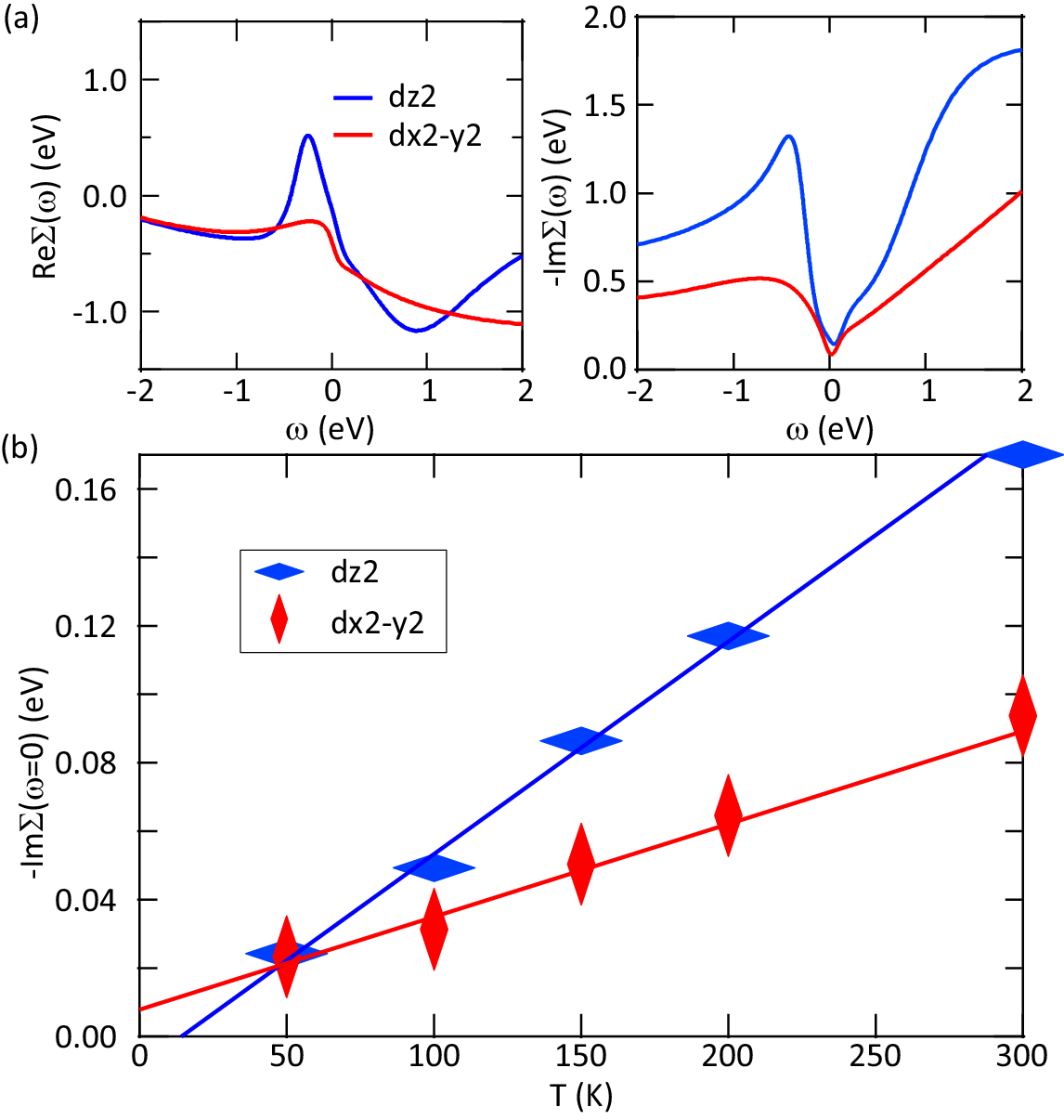}
        \caption{(a) Typical results for the real and imaginary parts of the orbital-dependent self-energy for $d_{z^2} $ (blue) and $d_{x^2-y^2}$ (red) orbitals in real frequency obtained using the maximum entropy method. (b) Temperature dependence of the imaginary part of the self-energy at zero frequency, -Im$\Sigma(\omega=0)$, calculated using DFT+DMFT for $U=5$ eV and $J=1$ eV. The solid line is a linear-in-temperature fit, showing strange metal behavior as observed in the resistivity measurement above the superconducting transition temperature.}
        \label{fig4}
    \end{center}
\end{figure}

Taken together, there appears to be an intimate interplay of orbital-selective Mott localization, Kondo hybridization, and Hund physics in superconducting La$_3$Ni$_2$O$_7$. The dual property of Ni-$d_{z^2}$ electrons is a characteristic feature of strongly correlated orbitals in Mott or Kondo systems. Since the $d_{z^2}$ orbitals are perpendicular to the $ab$ plane, their in-plane hopping is relatively small. Thus for the double-layer La$_3$Ni$_2$O$_7$, one may consider to start from Mott-localized $d_{z^2}$ electrons with a large inter-layer superexchange interaction, and explore their delocalization by hopping to neighboring Ni-$d_{x^2-y^2}$ orbitals. Their hybridization has two effects. It may not only create hybridization bands ($\Gamma$-X) near the Fermi energy as in Kondo lattice systems, but also produce self-doping to the localized $d_{z^2}$ orbitals as in hole-doped cuprates, causing $d_{z^2}$ quasiparticle band around M whose position depends on the interlayer hopping. The Hund's rule coupling may compete with the hybridization and greatly promote the band flatness. These results are distinctively different from the DFT weak correlation picture with almost fully-occupied $d_{z^2}$ bonding band. The possibility of Mott physics with strong renormalization has been confirmed in latest optical measurements at ambient pressure \cite{LiuZ2023}. 

Our strong correlation picture derived from the above DFT+DMFT calculations have important implications for understanding two major experimental observations in pressurized La$_3$Ni$_2$O$_7$, namely its strange metallicity in the normal state and its high-temperature superconductivity below around 80 K. To see how the linear-in-temperature resistivity may arise in our picture, we show in Fig. \ref{fig4} the self-energy $\Sigma(\omega)$ for both $d_{z^2}$ and $d_{x^2-y^2}$ orbitals in real frequency. The imaginary part of the self-energy at zero frequency gives the inverse of corresponding quasiparticle lifetimes at the Fermi energy. Their temperature dependencies are given in Fig. \ref{fig4}(b) up to 300 K. We find both can be well fitted using the linear-in-temperature function (solid lines). This gives a quasiparticle lifetime inversely proportional to the temperature, which is often used as an indicator of strange metallicity in the literature \cite{Hartnoll2022RMP}. Our results suggest that the strange metal behavior observed in La$_3$Ni$_2$O$_7$ comes from the flat quasiparticle bands around the Fermi energy due to strong electronic correlations.

Naturally, superconductivity born out of this strange metallic normal state should also originate from the pairing of strongly correlated $d_{z^2}$ and $d_{x^2-y^2}$ electrons, potentially mediated by antiferromagnetic spin fluctuations as revealed in our susceptibility calculations. It is possible that the large antiferromagnetic superexchange coupling between Ni-$d_{z^2}$ electrons may first lead to interlayer preformed pairs \cite{Keimer2015N, Bozovic2020NP}, which eventually give rise to the high-temperature superconductivity when phase coherence is established as the hybridization with the $d_{x^2-y^2}$ orbitals increases. This scenario is supported by our preliminary quantum Monte Carlo simulations for the above strongly correlated model \cite{Qin2023}. It will be interesting to see if pseudogap might also exist in this system.

To summarize, we have performed systematic DFT+DMFT studies of the strongly correlated electronic band structures of superconducting La$_3$Ni$_2$O$_7$ under high pressure. Our analyses reveal some key features in the electronic and magnetic properties, including the almost localized $d_{z^2}$ electrons, the very flat $d_{z^2}$ quasiparticle bands, and the strongly renormalized $d_{x^2-y^2}$ hybridization bands. We explain the normal state strange metal behavior from the linear-in-temperature dependence of the inverse quasiparticle lifetimes and find strong antiferromagnetic spin correlations that may be responsible for their Cooper pairing. Our observation of the flat band features, driven by the interplay of orbital-selective Mott, Hund, and Kondo physics, confirms the peculiarity of the double-layer La$_3$Ni$_2$O$_7$, and establishes a strong correlation picture for its possible high-temperature superconductivity. The coexistence of these competing mechanisms may give rise to even richer correlated many-body phenomena to be explored in future theoretical and experimental studies.

This work was supported by the National Natural Science Foundation of China (Grants No. 11974397, No. 12174429) and the Strategic Priority Research Program of the Chinese Academy of Sciences (Grant No. XDB33010100). We thank Guang-Ming Zhang for discussions and the Tianhe platforms at the National Supercomputer Center in Tianjin for technical support.

\end{document}